\documentclass[aps,pra,twocolumn]{revtex4}

\usepackage{graphics}
\usepackage{amsmath}
\usepackage{amssymb}
\usepackage{epsfig}
\usepackage{color}
\usepackage{caption}
\usepackage{subcaption}
\usepackage{bbm}	
\usepackage{adjustbox}
\graphicspath{{.}{./EPS/}}

\newcommand* {\bra}[1]{\ensuremath{\langle {#1} |}}
\newcommand* {\ket}[1]{\ensuremath{| {#1} \rangle}}

\begin{document}
 
\title{An entropy production based method for determining the position diffusion's coefficient of a quantum Brownian motion}

\author{J. Z. Bern\'ad}
\email{zsolt.bernad@um.edu.mt}
\affiliation{Department of Physics, University of Malta, Msida MSD 2080, Malta}
\affiliation{Institut f\"{u}r Angewandte Physik, Technische Universit\"{a}t Darmstadt, D-64289 Darmstadt, Germany}
\author{G. Homa}
\email{ggg.maxwell1@gmail.com}
\affiliation{Department of Physics of Complex Systems, E\"{o}tv\"{o}s Lor\'and University, ELTE, P\'azm\'any P\'eter s\'et\'any 1/A, H-1117 Budapest, Hungary}
\author{M. A. Csirik}
\email{csirik.mihaly@wigner.mta.hu}
\affiliation{Institute for Solid State Physics and Optics, Wigner Research Centre, Hungarian Academy of Sciences, 
P.O. Box 49, H-1525
Budapest, Hungary}

\date{\today}

%\author{G. Homa}
%\email{ggg.maxwell1@gmail.com}
%\affiliation{Department of Physics of Complex Systems, E\"{o}tv\"{o}s Lor\'and University, ELTE, P\'azm\'any P\'eter s\'et\'any 1/A, H-1117 Budapest, Hungary}
%\author{M. A. Csirik}
%\email{csirik.mihaly@wigner.mta.hu}
%\affiliation{Institute for Solid State Physics and Optics, Wigner Research Centre, Hungarian Academy of Sciences, 
%P.O. Box 49, H-1525
%Budapest, Hungary}

\date{\today}

\begin{abstract}
Quantum Brownian motion of a harmonic oscillator in the Markovian approximation is described by the respective Caldeira-Leggett master equation. 
This master equation can be brought into Lindblad form by adding a position diffusion term to it. The coefficient of this
term is either customarily taken to be the lower bound dictated by the Dekker inequality or determined by more detailed derivations
on the linearly damped quantum harmonic oscillator. In this paper, we explore the theoretical possibilities of 
determining the position diffusion term's coefficient by analyzing the entropy production of the master equation.
\end{abstract}

\maketitle

\section{Introduction}
\label{I}

Origins of the quantum Brownian motion lie in the question of how to reconcile damped equation of motion, e.g., Langevin equation for 
Brownian motion, with the process of quantization \cite{CLmodel}. Over the past decades, quantum dissipative systems have received 
a lot of attention \cite{Weiss} and derivations with respect to the quantum Brownian motion's master equation have been thoroughly scrutinized, see 
for example Refs. \cite{Grabert,Unruh,Hu,Fleming} (the literature on the topic is considerable). In the case of the Markovian limit, the master equation is expected to be in 
Lindblad form \cite{Gorini,Lindblad} in order to generate a completely positive trace preserving (CPTP) semigroup. However,
the so-called Caldeira-Leggett master equation of Ref. \cite{CLmodel}, which is based on functional integral description of damped
quantum systems \cite{Feynman}, is found not to be in Lindblad form \cite{LajosEu,Halliwell}. Another approach to the quantum mechanics of a 
damped harmonic oscillator \cite{Senitzky,Dekker1,Dekker2}, where a weak coupling is considered between the harmonic oscillator and the reservoir,
has resulted in a similar master equation to Ref. \cite{CLmodel}, but for this time in Lindblad form. The master equation in \cite{CLmodel} 
with momentum diffusion and damping terms is now extended with two diffusion terms, where one of them describes position diffusion and the other
term is a double commutator involving both position and momentum operator. The coefficients of the four diffusion terms must satisfy the 
Dekker inequality \cite{Dekker2} in order that the master equation generates a CPTP semigroup. However,
if one wanted to determine the value of one of the coefficients, provided that the other three coefficients are known, the Dekker inequality would 
result only in a lower bound. Therefore, in the context of bringing the Caldeira-Leggett master equation in Lindblad form, the coefficient of the 
position diffusion term is not exactly determined which is a curious fact and deserves attention. The other three coefficients of the diffusion terms 
can be obtained in the Markovian limit, see for example \cite{Hu}. A possible approach to this question is to 
reconsider the approximations within the microscopic model and derive a new master equation. Ref. \cite{Diosi1993} 
has extended the derivation of Ref. \cite{CLmodel} to medium temperatures, thus obtaining all coefficients by neglecting quadratic and 
higher orders in inverse temperature. In a different approach, H. Dekker has obtained these coefficients by applying a quasi-canonical phase 
space quantization procedure to the linearly damped harmonic oscillator \cite{Dekker3}. 

The purpose of this paper is to present a method, which is also capable of determining the value of the position diffusion's coefficient. 
We consider the master equation of Ref. \cite{CLmodel} and extend it such that it is in Lindblad form and by thus assuring the mathematical consistency. Three coefficients out of four of the
diffusion terms are obtained by assuming that the environment has Ohmic spectral density with a high frequency cutoff. In fact, we use most of the arguments of Ref. \cite{Diosi1993} and only neglect 
the result on the coefficient of the position diffusion term, which we consider to be unknown. We simply pose the question: Is it possible to determine this value
only from the mathematical structure of the master equation? We are going to investigate this question with the help of the entropy production \cite{Spohn}. The entropy production
is defined for quantum dynamical semigroups and this is why the more general non-Markovian master equations of the Caldeira-Leggett model are excluded from our approach.

The generator of the extended Caldeira-Leggett master equation is an unbounded operator and it is important to note this, because the proof 
of Lindblad \cite{Lindblad} is based on uniformly continuous semigroups. Therefore, we use a Gaussian ansatz for the density matrix, 
which is in the domain of the generator, and furthermore, states characterized by this Gaussian ansatz preserve their structure during the whole time evolution. 
In this context, the resulting density matrix equations enable investigations on the relative entropy between any evolving state and 
the steady state of the master equation. Now, taking the negative time derivative of this special relative entropy functional at $t=0$, we get the
entropy production \cite{Spohn}. As the entropy production is a convex functional on the state space and vanishes in the steady state of the system, one is able to 
consider the principle of minimal or maximal entropy production under present constraints \cite{Prigogine} (for an extended view see Ref. \cite{Martyushev}). As the formulation
of the minimal entropy production is strictly valid only in the neighborhood of equilibrium \cite{Prigogine} and the steady state of the extended Caldeira-Leggett master equation 
is not guaranteed to be in the thermal equilibrium, we are going to search for both maximum and
minimum of the entropy production. We determine the coefficient of the position diffusion term from these extrema under the constraint of the Dekker inequality.
In our case the entropy production is going to depend on the initial conditions of the system from which it evolves towards the steady state. Therefore,
we are going to analyze several type of initial conditions, involving also a case where the initial condition is in the close neighborhood of the steady state.  

The paper is organized as follows. In Sec. \ref{II} we discuss the solutions of the master equation and give the analytical formula
of the relative entropy between an arbitrary state characterized by our Gaussian ansatz and the steady state. In Sec. \ref{III} we 
derive the formula for the entropy production rate and explore its properties. We determine the coefficient of the position diffusion term
for different initial conditions. Detailed derivations supporting the main text are collected in two appendices.

\section{Master equation and relative entropy}
\label{II}

Let us consider the extended Markovian Caldeira-Leggett master equation \cite{CLmodel,Hu,Dekker1} of a quantum harmonic oscillator with mass $m$ and 
frequency $\omega$
($\hbar=1$)
\begin{eqnarray}
\frac{d\hat{\rho}}{dt}&=&-i\left [\frac{\hat{p}^2}{2m}+\frac{m \omega^2 \hat{x}^2}{2},\hat{\rho}\right]-D_{pp}[\hat{x},[\hat{x},\hat{\rho}]] -i \gamma [\hat{x},\{\hat{p},\hat{\rho}\}]\nonumber\\
&&+2D_{px}[\hat{x},[\hat{p},\hat{\rho}]]-D_{xx}[\hat{p},[\hat{p},\hat{\rho}]],\label{ME}
\end{eqnarray}
subject also to a position diffusion with coefficient $D_{xx}$ and where $[,]$ stands for commutators while $\{,\}$ for the anti-commutators. 
$\gamma$ is the relaxation constant and $D_{pp}$ stands for the momentum diffusion coefficient. $D_{px}$ is a cross diffusion coefficient. The above master equation
without the term $[\hat{p},[\hat{p},\hat{\rho}]]$ is derived from the Caldeira-Leggett model \cite{CLmodel} by considering an environment of harmonic oscillators in 
thermal equilibrium with temperature $T$ and with Ohmic spectral density and a high frequency cutoff $\Omega$ (see e.g. Refs. \cite{Breuer}). Furthermore, 
the Born-Markov approximation, i.e., $\Omega, k_B T\gg \gamma$, is employed during the derivation and the slow motion of the central
system compared to the bath correlation time, i.e., $\Omega, k_B T\gg \omega$, is also considered. The extra term 
$[\hat{p},[\hat{p},\hat{\rho}]]$ has already been derived in other approaches, for instance, in the pioneering stage of investigating
quantum dissipative systems \cite{Dekker1}, or later as an extension of the Caldeira-Leggett model to medium temperatures \cite{Diosi1993}.

In the next step we
consider a representation of the density matrix in \eqref{ME} with the help of a double Fourier transform \cite{Unruh}
\begin{equation}
\rho(k,\Delta,t)=\mathrm{Tr} \big\{\hat{\rho}(t) \exp\{ik \hat{x}+ i \Delta \hat{p}\} \big\}. \label{transform}
\end{equation}
The equation of motion for $\rho(k,\Delta,t)$ reads
\begin{eqnarray}
 \frac{\partial}{\partial t} \rho(k,\Delta,t)&=&\Big (\frac{k}{m} \frac{\partial}{\partial \Delta} 
 -m \omega^2 \Delta \frac{\partial}{\partial k}-D_{pp} \Delta^2-2\gamma \Delta \frac{\partial}{\partial \Delta}
 \Big. \nonumber \\
&& \Big. -2D_{px}k \Delta-D_{xx} k^2 \Big) \rho(k,\Delta,t), \nonumber
\end{eqnarray}
where we used the relations
\begin{eqnarray}
 e^{ik \hat{x}+ i \Delta \hat{p}} \hat{p}&=&-\Big(i\frac{\partial}{\partial \Delta} +\frac{k}{2}\Big) e^{ik \hat{x}+ i \Delta \hat{p}}, \nonumber\\
 \hat{p} e^{ik \hat{x}+ i \Delta \hat{p}}&=&-\Big(i\frac{\partial}{\partial \Delta} -\frac{k}{2}\Big) e^{ik \hat{x}+ i \Delta \hat{p}}
 \nonumber \\
 \hat{x} e^{ik \hat{x}+ i \Delta \hat{p}}&=&-\Big(i\frac{\partial}{\partial k} +\frac{\Delta}{2}\Big) e^{ik \hat{x}+ i \Delta \hat{p}}, \nonumber\\
   e^{ik \hat{x}+ i \Delta \hat{p}}\hat{x}&=&-\Big(i\frac{\partial}{\partial k} -\frac{\Delta}{2}\Big) e^{ik \hat{x}+ i \Delta \hat{p}}.
 \nonumber 
\end{eqnarray}
In order to solve this equation of motion we make use of the following Gaussian ansatz
\begin{eqnarray}
\rho(k,\Delta,t)&=&\exp\{-c_1(t)k^2-c_2(t)k \Delta -c_3(t) \Delta^2-ic_4(t) k \nonumber\\
&& -ic_5(t) \Delta-c_6(t)\}, \label{Ga}
\end{eqnarray}
where the parameters $c_1,c_2,c_3,c_4,c_5$ and $c_6$ are real and they obey the following linear differential equations
\begin{eqnarray}
 \dot{c}_1&=&D_{xx}+\frac{c_2}{m}, \quad \dot{c}_2=2D_{px}+\frac{2c_3}{m}-2 m \omega^2 c_1-2\gamma c_2,
 \nonumber \\
 \dot{c}_3&=&D_{pp}-m \omega^2 c_2-4\gamma c_3, \quad 
 \dot{c}_4=\frac{c_5}{m}, \nonumber \\
 \dot{c}_5&=&-m \omega^2 c_4-2\gamma c_5, \quad \dot{c}_6=0. \label{difftosolve}
\end{eqnarray}
These differential equations can be solved by standard methods (see Appendix \ref{AppI}), but the general solutions are not required for our study.
We only present here the steady state solutions (see also \eqref{stacsol}):
\begin{eqnarray}
 c^{\text{st}}_1&=&\frac{4\gamma m D_{px}+m^2 \left(4 \gamma ^2+\omega ^2\right) D_{xx}+D_{pp}}{4 \gamma  m^2 \omega ^2},
 \nonumber\\
  c^{\text{st}}_2&=&-m D_{xx}, \quad
 c^{\text{st}}_3=  \frac{m^2 \omega ^2 D_{xx}+D_{pp}}{4 \gamma }, \nonumber \\  
 c^{\text{st}}_4&=&c^{\text{st}}_5=0, \quad c^{\text{st}}_6=c_6(0). \label{steadyci} 
\end{eqnarray}
The eigenvalues of the density matrix $\hat{\rho}$ with the Gaussian ansatz \eqref{Ga} in the $(k,\Delta)$ representation 
are obtained in the equivalent position representation (see Appendices \ref{AppI} and \ref{AppII}) 
\begin{eqnarray}
\rho(x,y,t)&=&\exp\{-A (t)  \left( x-y \right)^2-iB(t)  \left( x-y \right)  \left( x+y \right) \nonumber\\
&&-C(t) \left( x+y \right)^{2}-iD(t) (x-y)
\nonumber \\
& &-E(t)(x+y) -N(t)\}, \label{xyform}
\end{eqnarray}
where all the time-dependent parameters $A(t)$, $B(t)$, $C(t)$, $D(t)$, $E(t)$ and $N(t)$ are real and \eqref{xyform} evolves according to
\begin{eqnarray}
i \frac{\partial}{\partial t} \rho(x,y,t)&=&\Big[ \frac{1 }{2 m }\left(\frac{\partial^2}{\partial y^2}-
\frac{\partial^2}{\partial x^2} \right) +\frac{m \omega^2}{2} \left(x^2-y^2\right) \Big. \nonumber\\
&&\Big.-i D_{pp} (x-y)^2-i \gamma (x-y) \left(\frac{\partial}{\partial x}-
\frac{\partial}{\partial y} \right) \nonumber \\
&&+2D_{px}(x-y) \left(\frac{\partial}{\partial x}+
\frac{\partial}{\partial y} \right)  \nonumber\\
&& \Big.+i D_{xx} \left(\frac{\partial}{\partial x}-
\frac{\partial}{\partial y} \right)^2 \Big]  \rho(x,y,t). \nonumber
\end{eqnarray}
Applying the transformation \eqref{BtoS} between the coefficients of $\rho(x,y)$ and $\rho(k,\Delta)$ the eigenvalues yield
\begin{equation}
 \lambda_n= \frac{2}{2\sqrt{4c_1c_3-c^2_2}+1} \left(\frac{2\sqrt{4c_1c_3-c^2_2}-1}{2\sqrt{4c_1c_3-c^2_2}+1}\right)^n, \quad n \in \mathbb{N}_0.
\end{equation}
Recall that thermal states of quantum harmonic oscillators in the Fock representation have the form
\begin{equation}
 \hat{\rho}_{\text{th}}=\frac{1}{\bar n +1}\sum^\infty_{n=0} \left( \frac{\bar n}{\bar n+1} \right)^n \ket{n}\bra{n},
\end{equation}
where $\bar n$ is the mean excitation number. We can already conclude that the spectrum of a thermal state is structurally identical with the spectrum of 
$\hat{\rho}$ with the Gaussian ansatz in \eqref{xyform}. Through the following simple identification
\begin{equation}
 \bar n= \frac{2\sqrt{4c_1c_3-c^2_2}-1}{2} \label{nbar}
\end{equation}
we see that $\hat{\rho}$ is a unitarily transformed thermal state with $(2\sqrt{4c_1c_3-c^2_2}-1)/2$ 
mean excitation number. Furthermore, $\rho(k,\Delta)$ is also 
the symmetric characteristic function $\chi(\lambda,\lambda^*)$ of $\hat{\rho}$
\begin{eqnarray}
\rho(k,\Delta)&=&\mathrm{Tr} \big\{\hat{\rho} \exp\{ik \hat{x}+ i \Delta \hat{p}\} \big\}=
\mathrm{Tr} \big\{\hat{\rho} \exp\{\lambda \hat{a}^\dagger- \lambda^* \hat{a}\} \big\} \nonumber\\
&=&\chi(\lambda,\lambda^*), \nonumber \\
k&=&\frac{-i}{\sqrt{2}} (\lambda-\lambda^*), \quad \Delta=\frac{-1}{\sqrt{2}} (\lambda+\lambda^*),
\end{eqnarray}
where $\hat{a}$ ($\hat{a}^\dagger$) is the annihilation (creation) operator. A simple transformation yields
\begin{eqnarray}
 \chi(\lambda,\lambda^*)= \exp \Big\{&-&(c_1+c_3)|\lambda|^2-\frac{-c_1+c_3+ic_2}{2}\lambda^2
 \Big.\nonumber \\
 &-&\frac{-c_1+c_3-ic_2}{2}(\lambda^*)^2 
 +\frac{-c_4+ic_5}{\sqrt{2}}\lambda \nonumber \\
 &-&\Big. \frac{-c_4-ic_5}{\sqrt{2}}\lambda^*-c_6 \Big\},
\end{eqnarray}
which finally shows that $\hat{\rho}$ with the Gaussian ansatz in the $(k,\Delta)$ representation is a type of displaced squeezed thermal state (DSTS) \cite{Marian}. 
Nevertheless, due to the evolution in \eqref{ME} and by choosing an initial Gaussian density matrix, $\hat{\rho}$ remains
a DSTS at all times.  

Our next aim is to calculate the relative entropy between $\hat{\rho}(t)$ and the stationary solution $\hat{\rho}^{\text{st}}$ for any 
$t \geqslant 0$. The relative entropy between the two arbitrary states $\hat{\rho}$ and $\hat{\sigma}$ is defined as \cite{PetzOhya}
\begin{equation}
 S(\hat{\rho} \,|\, \hat{\sigma})=\begin{cases} \mathrm{Tr} \{\hat{\rho} (\log \hat{\rho}-\log \hat{\sigma})\}, 
 & \mathrm{supp}(\rho) \subseteq \mathrm{supp}(\sigma) \\ +\infty, & \text{otherwise}. \end{cases} \label{relentrsc}
\end{equation}
The support of a state $\hat{\rho}$ is the complement of its kernel, i.e., a subspace of the Hilbert space where $\hat{\rho}$ does not have
eigenvalues equal to zero. 

According to Ref. \cite{Marian} the relative entropy between two DSTS states has an analytical form and can be expressed as a function of the
two characteristic functions' parameters. Employing this formula we obtain the following relation:
\begin{eqnarray}
 &&S(\hat{\rho}(t) \,|\, \hat{\rho}^{\text{st}})=-(\bar n(t)+1) \log (\bar n(t)+1)+ \bar n(t) \log \bar n(t)\nonumber \\
 +
&& \frac{\log\left[(\bar n^{\text{st}}+1)\bar n^{\text{st}}\right]}{2} 
+\frac{2 \log \left(\frac{\bar n^{\text{st}}+1}{\bar n^{\text{st}}}\right)}{2\bar n^{\text{st}}+1} \Big[ 2 c_1(t)c^{\text{st}}_3 
 +2c_3(t)c^{\text{st}}_1 \Big.\nonumber \\
  && \Bigl. -c_2(t) c^{\text{st}}_2+c^{\text{st}}_1 c^2_5(t)+c^{\text{st}}_3 c^2_4(t)
-c_4(t)c_5(t)c^{\text{st}}_2 \Big], \label{entropyan}
\end{eqnarray}
where we have used the relation \eqref{nbar} to define 
\begin{eqnarray}
 \bar n(t)&=& \frac{2\sqrt{4c_1(t)c_3(t)-c^2_2(t)}-1}{2},
 \nonumber\\
  \bar n^{\text{st}}&=& \frac{2\sqrt{4c^{\text{st}}_1c^{\text{st}}_3-
 (c^{\text{st}}_2)^2}-1}{2}.  
\end{eqnarray}

We remark that \eqref{entropyan}, cast in an analytical form, is better suited for our subsequent investigation.

\section{Entropy production and position diffusion}
\label{III}

In this section we present and investigate an entropy production based method, which is capable of identifying the coefficient of 
the position diffusion term in \eqref{ME}.  We consider the high temperature limit $k_B T\geqslant \Omega \gg \omega$ in order
to obtain the coefficients of the well-known Caldeira-Leggett master equation \cite{CLmodel}, thus yielding the
following relations
\begin{equation}
 \gamma = \frac{\eta}{2m},\quad D_{pp}=\eta k_B T, \quad D_{px}=\frac{\gamma k_B T}{\Omega} \label{derivedcoeff},
\end{equation}
where $\eta$ is the viscosity coefficient in a Quantum Brownian motion. We immediately have
that $D_{pp}=2m \gamma k_B T$. These coefficients can also be determined from 
an extension of the Caldeira-Leggett master equation to medium temperatures (see Ref. \cite{Diosi1993}).

The Markovian master equation in \eqref{ME} is not in the Lindblad form \cite{Gorini,Lindblad}, which presents a mathematical difficulty since the 
Lindblad form would ensure that the master equation generates a quantum dynamical or CPTP semigroup. This semigroup maps quantum states to 
quantum states and furthermore the map is also completely positive. We shall not discuss here the properties and applications 
of quantum dynamical semigroups, instead we refer to the book \cite{Breuer}. So the Caldeira-Leggett model can be put in Lindblad form by 
adding $-D_{xx}[\hat{p},[\hat{p},\hat{\rho}]]$ and we already know that $D_{xx}$ can be defined through the 
Dekker inequality \cite{Dekker2,SaSc}
\begin{equation}
 D_{pp}D_{xx}-D^2_{px}\geqslant \frac{\gamma^2}{4}, \label{Dekkerineq}
\end{equation}
obtained by imposing the condition that Eq. \eqref{ME} preserves the uncertainty principle \cite{Dekker2} or due to the algebraic relation of the coefficients
in the Lindblad form \cite{SaSc}. Combining this with
\eqref{derivedcoeff} we find that the minimal value for $D_{xx}$ is
\begin{equation}
 D^{\text{min}}_{xx}=\frac{\gamma}{8mk_BT}+\frac{\gamma k_BT}{2m \Omega^2}. \label{minvalue}
\end{equation}
Inequality \eqref{Dekkerineq} is also a fundamental constraint, which must be satisfied in order to write the master equation \eqref{ME} 
in Lindblad form. In Ref. \cite{Diosi1993}, a value of $D_{xx}=\gamma/(6mk_BT)$ was obtained for large and 
medium temperatures by using Markovian approximation in a path integral formalism. 
This method neglected the quadratic and higher powers of the inverse temperature $1/T$.

\begin{figure}
\centering
\begin{subfigure}{.5\textwidth}
  \centering
  \includegraphics[width=0.9\linewidth]{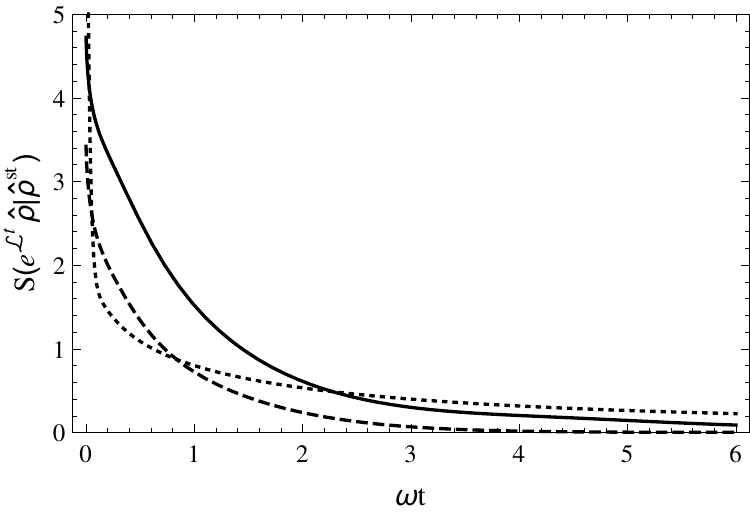}
  \caption{ Coherent state $\ket{\alpha}$ with $\alpha=2 + 2 \cdot i$.}
  \label{fig1}
\end{subfigure}%
\\
\begin{subfigure}{.5\textwidth}
  \centering
  \includegraphics[width=0.9\linewidth]{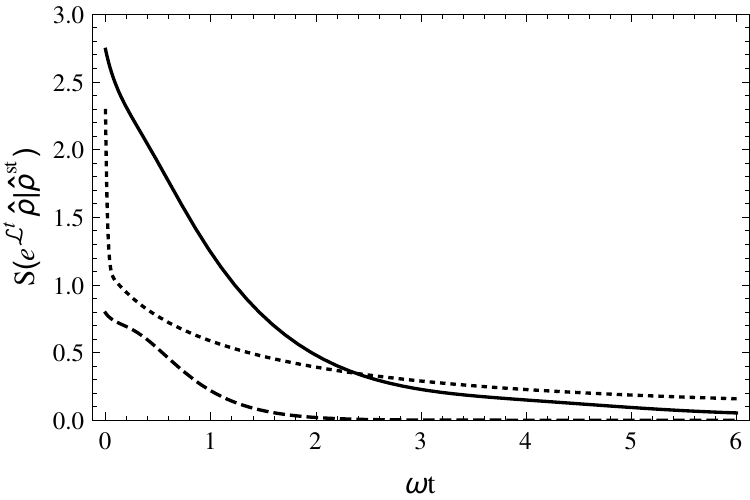}
  \caption{ Thermal state with $\bar n=2$.}
  \label{fig2}
\end{subfigure}
\caption{Relative entropy $S\big(e^{\mathcal L t} 
 \hat{\rho} \,|\,\hat{\rho}^{\text{st}}\big)$ as a function of time according to Eq. \eqref{deff}. The time evolution is presented for 
 two different initial conditions, a coherent state and a thermal state. We set $D_{xx}$ to its minimum allowed value,
 defined by the inequality \eqref{Dekkerineq}, the dimensionless parameter $D_{pp}/(2m\omega^2)=10$, and $\omega/\Omega=0.1$. In both figures three type of
 curves are shown for different values of $\gamma/\omega$: $0.1$ (solid), $1$ (dashed), and $10$ (dotted). It is worth to note that for different values of
 $\gamma/\omega$ the system approaches a different steady state, see Eq. \eqref{steadyci}.} \label{fig12}
\end{figure}

In order to find possible values for $D_{xx}$ satisfying inequality \eqref{Dekkerineq}, we consider a different method. It has been shown
by Refs. \cite{Lindblad2,Uhlmann} that for a CPTP map $\Phi$,
\begin{equation}
 S(\hat{\rho} \,|\, \hat{\sigma}) \geqslant S\big(\Phi(\hat{\rho}) \,|\, \Phi(\hat{\sigma})\big), \label{mono}
\end{equation}
i.e. the monotonicity of the quantum relative entropy holds. Now, rewriting Eq. \eqref{ME} as
\begin{equation}
 \frac{d\hat{\rho}}{dt}=\mathcal{L} \hat{\rho}, \quad \text{with} \quad \Phi=e^{\mathcal L t}, \nonumber 
\end{equation}
we have that the time-dependent function
\begin{equation}
 S\big(e^{\mathcal L t} \hat{\rho}(0) \,|\, e^{\mathcal L t} \hat{\rho}^{\text{st}}\big)=S\big(e^{\mathcal L t} 
 \hat{\rho}(0) \,|\,\hat{\rho}^{\text{st}}\big)
 \label{deff} 
\end{equation}
is monotonically decreasing in time and continuous from the right (see Ref. \cite{Spohn}). This time dependence is depicted in Figs. \ref{fig1} and \ref{fig2}
for two types of initial conditions, a coherent state and a thermal state. For the underdamped $\gamma/\omega<1$ and overdamped $\gamma/\omega>1$
cases the relative entropy approaches 0 slower than in the critically damped case $\gamma/\omega=1$. This is due to 
the characteristic frequencies present in the evolution $e^{\mathcal L t} \hat{\rho}$ (see Appendix \ref{AppI}): 
$-2\gamma/\omega \pm 2\sqrt{(\gamma/\omega)^2-1}$, $-\gamma/\omega \pm \sqrt{(\gamma/\omega)^2-1}$, and $-2\gamma$. 
When we have an underdamped case then the above complex frequencies have small real parts, thus the relative entropy approaches zero slowly. 
In the overdamped case, the frequencies  $-\gamma/\omega + \sqrt{(\gamma/\omega)^2-1}$ and 
 $-2\gamma/\omega + 2\sqrt{(\gamma/\omega)^2-1}$ are again small, which results in a slow exponential decay.   
 
The entropy production is defined as the negative time-derivative of \eqref{deff} at $t=0$,
\begin{equation}
 \sigma=-\frac{d}{dt} S\big(e^{\mathcal L t} \hat{\rho}(0) \,|\,\hat{\rho}^{\text{st}}\big) \Big | _{t=0}, \label{entrprod}
\end{equation}
which is a nonnegative convex functional. In the following we conduct an analysis based on the extremum of $\sigma$ as a function of $D_{xx}$, a condition, which 
allows us to acquire possible values for $D_{xx}$.

%%%%%%%%%%%%%%%%%%%%%%%%%%%%%%%%%%%%%%%%%%%%%%%%%%%%%%%%%%%%%%%%%%%%%%%%%%%%%%%
%%%%%%%%%%%%%%%%%%%%%%%%%%%%%%%%%%%%%%%%%%%%%%%%%%%%%%%%%%%%%%%%%%%%%%%%%%%%%%%%%%

In order to investigate the entropy production rate we express $\sigma$ as the function of $\gamma$, $D_{pp}$, $D_{px}$ and $D_{xx}$. First, we substitute
the relations
\begin{eqnarray}
\dot{c}_1(0)&=&D_{xx}+\frac{c_2(0)}{m}, \nonumber \\ \dot{c}_2(0)&=&2D_{px}+
 \frac{2c_3(0)}{m}-2 m \omega^2 c_1(0)-2\gamma c_2(0), \nonumber \\
\dot{c}_3(0)&=&D_{pp}-m \omega^2 c_2(0)-4\gamma c_3(0), \quad
\dot{c}_4(0)=\frac{c_5(0)}{m}, \nonumber \\
\dot{c}_5(0)&=&-m \omega^2 c_4(0)-2\gamma c_5(0) \nonumber
\end{eqnarray}
into \eqref{entrprod}, where we make use of Eqs. \eqref{steadyci} and \eqref{entropyan}. We finally find that
\begin{eqnarray}
&&\sigma/\omega=\nonumber \\ && \frac{4 D'_{pp} c'_1(0)-4D'_{px} c'_2(0) +4 D'_{xx} c'_3(0)-\gamma' \big(2 \bar n(0)+1\big)^2}{2 \bar n(0)+1}  
 \label{sigmaan} \nonumber  \\
&&\times \log \left(\frac{\bar n(0)+1}{\bar n(0)}\right)-\log \left(\frac{\bar n^{\text{st}}+1}{\bar n^{\text{st}}}\right) \Bigg[4 D'^2_{pp} + \frac{D'^2_{xx}}{4} \Bigg. \nonumber \\  &&\Bigg.
 +2 D'_{pp} D'_{xx}  (2 \gamma'^2+1)- 8 \gamma' D'_{pp} \big(2 c'_3(0)+c'^2_5(0)\big) \Big. \nonumber \\
 \Big.
&&+2 \gamma' D'_{px} \Big(4 D'_{pp}+D'_{xx} -2 c'_2(0) -2 c'_4(0) c'_5(0)-16 \gamma' c'_3(0) \Bigg.\nonumber \\ \Bigg.
&&-8 \gamma' c'^2_5(0) \Big)-\gamma' D'_{xx} \Bigg(c'_1(0)+4 \gamma' c'_2(0)+16 \gamma'^2 c'_3(0) \nonumber \\ && +\frac{\big(c'_4(0)+4\gamma' c'_5(0)\big)^2}{2}\Bigg)\Bigg]  \frac{1}{\gamma'(2  \bar n^{\text{st}} +1)},
\end{eqnarray}
where we have introduced the following dimensionless parameters
\begin{eqnarray}
D'_{pp}&=&\frac{D_{pp} x^2_0}{\omega}, \gamma'=\frac{\gamma}{\omega},  D'_{xx}=\frac{D_{xx}}{\omega x^2_0},  D'_{px}=\frac{D_{px}}{\omega}, \label{transf} \\
 \quad c'_1&=&\frac{c_1}{x^2_0}, c'_2=c_2,  c'_3=c_3 x^2_0, c'_4=\frac{c_4}{x_0}, 
 c'_5=c_5 x_0, \nonumber 
\end{eqnarray}
and $x_0=1/\sqrt{2 m \omega}$ is the width of the quantum harmonic oscillator's ground state wave function. It is worth mentioning
 that
$\bar n^{\text{st}}$ and $\bar n(t)$ ($t\geqslant 0$) are invariant under the above transformations.

Now, we investigate those cases where the entropy production approaches infinity.
These cases are inconclusive and one cannot infer any information about $D'_{xx}$. We would like to keep all our parameters finite and therefore
it can be seen that there are two cases: whenever
either $\bar n(0)$ or $\bar n^{\text{st}}$ approaches zero the entropy production rate approaches infinity. These limit cases can be regarded as the 
system is either initially or stationarily in a pure state. The latter one is equivalent to
\begin{equation}
 4c^{\text{st}}_1c^{\text{st}}_3-(c^{\text{st}}_2)^2=\frac{1}{4} \label{mnbv}
\end{equation}
and substituting \eqref{steadyci} into it we obtain a quadratic equation for $D'_{xx}$, which has two real roots:
\begin{eqnarray}
 && D'^{\pm}_{xx}=-4 D'_{pp}\big(1+2\gamma'^2\big) -4 \gamma'D'_{px} \nonumber \\
&& \pm 2 \gamma' \sqrt{1+16 D'_{pp} \big(1+\gamma'^2 \big)+4 D'_{px} \big(D'_{px}+4 \gamma'D'_{pp}\big) }.\qquad \label{rootD}
\end{eqnarray}
Now, comparing the bigger root $D'^{+}_{xx}$ with the minimum of $D'_{xx}$ allowed by \eqref{Dekkerineq} we have
\begin{equation}
 D'^{\text{min}}_{xx}=\frac{\gamma'^2}{4 D'_{pp}}+\frac{D'^2_{px}}{D'_{pp}} \geqslant D'^{+}_{xx}, \nonumber
\end{equation}
which is equivalent to 
\begin{eqnarray}
&& \left(\gamma'^2-16 D'^2_{pp}\right)^2+ 64 \gamma'^4  D'^2_{pp}+8 D'_{px} \Big[64 \gamma' D'^3_{pp} \Big.
 \nonumber\\
&& +16\left(1+2\gamma'^2\right)D'^2_{pp}D'_{px} +D'_{px} \left(\gamma'^2+2D'^2_{px}\right) \nonumber \\
&& \Big. +4 D'_{pp} \left(\gamma'^3+4 \gamma' D'^2_{px}\right) \Big]  \geqslant 0. \nonumber
\end{eqnarray}
As $\gamma'$, $D'_{pp}$, and $D'_{px}$ are positive real numbers, the above expression is always true and the equality is obtained only for $\gamma'=D'_{pp}=0$, which is an uninteresting case 
for our study. 
Hence, from now on we shall investigate the global extrema of the entropy production $\sigma$ only on the interval $[D'^{\text{min}}_{xx}, \infty)$. $\sigma$ depends on the initial conditions 
of the system and 
therefore we are going to study some cases below. 

{\it Initial state is a displaced and squeezed steady state.} The initial condition reads
\begin{equation}
 c'_1(0)=c'^{\text{st}}_1, \quad c'_2(0)=c'^{\text{st}}_2,\quad c'_3(0)=c'^{\text{st}}_3, \label{condspec}
\end{equation}
with $c'_4(0)$ and $c'_5(0)$ having arbitrary values, which directly yields that $\bar n(0)=\bar n^{\text{st}}$. We also make the observation that
\begin{equation}
 \lim_{D'_{xx} \to \infty} \frac{\sigma}{\omega D'_{xx}}=\frac{4 c'_3(0)}{2 \bar n(0)+1} \log \left(\frac{\bar n(0)+1}{\bar n(0)}\right)=a, \label{a}
\end{equation}
and 
\begin{eqnarray}
 &&\lim_{D'_{xx} \to \infty} \frac{\sigma}{\omega}-a D'_{xx}
 = \nonumber \\ 
&& \frac{4 D'_{pp} c'_1(0)-4 D'_{px} c'_2(0)-\gamma' \big(2 \bar n(0)+1\big)^2}{2 \bar n(0)+1}  
  \log \left(\frac{\bar n(0)+1}{\bar n(0)}\right) 
  \nonumber \\
 && -2\gamma'=b,
\end{eqnarray}
which means that $\sigma/\omega$ as a function of $D'_{xx}$ is asymptotic to the line $a D'_{xx}+b$. The slope of the asymptotic line in Eq. \eqref{a} for the initial condition \eqref{condspec} reads
\begin{equation}
 \lim_{D'_{xx} \to \infty}\frac{4 c'_3(0)}{2 \bar n(0)+1} \log \left(\frac{\bar n(0)+1}{\bar n(0)}\right)=0.
\end{equation}

%%%%%%%%%%%%%%%%%%%%%%%%%%%%%%%%%%%%%%%%%%%%%%%%%%%%%%%%%%%%%%%%%%%%%%%%%%%%%%%
\begin{figure}
\centering
\begin{subfigure}{.5\textwidth}
  \centering
  \includegraphics[width=0.9\linewidth]{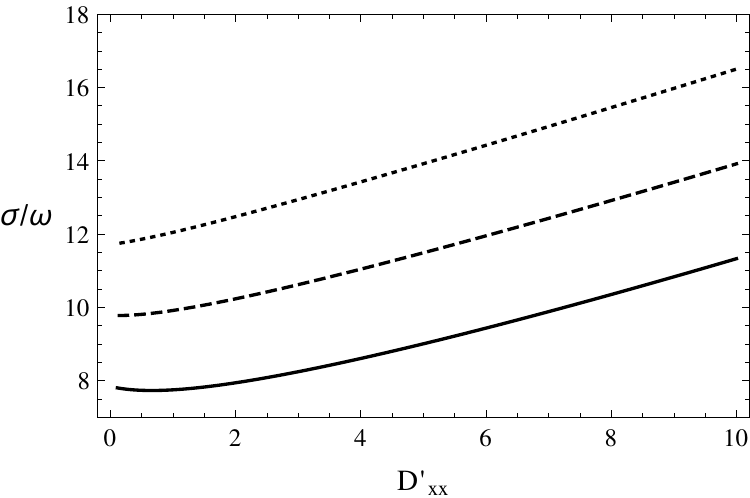}
  \caption{ Initial state with $\bar n(0) \neq \bar n^{\text{st}}$.}
  \label{fig3}
\end{subfigure}\\
\begin{subfigure}{.5\textwidth}
  \centering
  \includegraphics[width=0.9\linewidth]{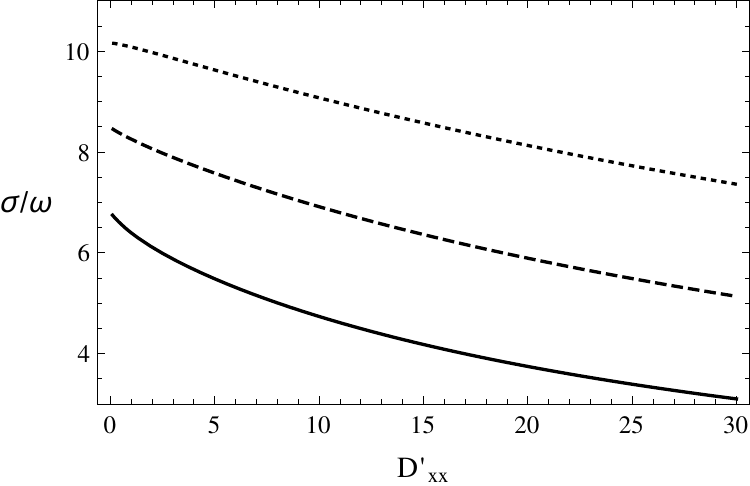}
  \caption{ Displaced and squeezed steady state.}
  \label{fig4}
\end{subfigure}
\caption{Entropy production $\sigma/\omega$ as a function of $D'_{xx}$ according to Eqs. \eqref{sigmaan} and \eqref{sigmax}. 
We set $c'_4(0)=c'_5(0)=2$, $D'_{pp}/\gamma'=2$, and $D'_{px}/\gamma'=0.2$ for all three type of curves: 
$\gamma'=0.8$ (solid), $\gamma'=1$ (dashed), and $\gamma'=1.2$ (dotted).
The figures are displayed for two different initial conditions: an initial state with $c'_1(0)=c'_2(0)=c'_3(0)=1$,  
which is unitarily not equivalent with the steady state; a displaced and squeezed steady state, i.e., $c'_1(0)=c'^{\text{st}}_1$,
$c'_2(0)=c'^{\text{st}}_2$, and $c'_3(0)=c'^{\text{st}}_3$. In both figures the curves start at the minimum allowed value 
in accordance with \eqref{Dekkerineq}.} \label{fig34}
\end{figure}
%%%%%%%%%%%%%%%%%%%%%%%%%%%%%%%%%%%%%%%%%%%%%%%%%%%%%%%%%%%%%%%%%%%%%%%%%%%%%%%%%%

Thus the global maximum of $\sigma$ at $D'_{xx} \to \infty$ turns into a global minimum, as it is shown in
Fig. \ref{fig34} and now $\sigma$ has a global maximum at a finite value of $D'_{xx}$ and varies as a function of $\gamma'$. The position
of the global minimum is physically uninteresting, which leaves us with the global maximum for determining the value of $D_{xx}$. In this situation the system turns back to its steady state 
as fast as possible. Finally, Eq. \eqref{sigmaan} under the conditions \eqref{condspec} is changed
to
\begin{eqnarray}
 \!\!\!&\!\!\!&\!\!\!\sigma\big(D'_{xx}\big)/\omega= \Big[16 D'_{pp} c'^2_5(0)+8 D'_{px} c'_5(0)\big(c'_4(0)+
 4\gamma' c'_5(0)\big)\nonumber \\
&&  + D'_{xx} \big(c'_4(0)+4\gamma' c'_5(0)\big)^2 \Big]  \frac{1}
 {4\bar n^{\text{st}}+2}
 \log \left(\frac{\bar n^{\text{st}}+1}{\bar n^{\text{st}}}\right). \label{sigmax}
\end{eqnarray}
%%%%%%%%%%%%%%%%%%%%%%%%%%%%%%%%%%%%%%%%%%%%%%%%%%%%%%%%%%%%%%%%%%%%%%%%%%%%%%%
\begin{figure}
\centering
\begin{subfigure}{.5\textwidth}
  \centering
  \includegraphics[width=0.9\linewidth]{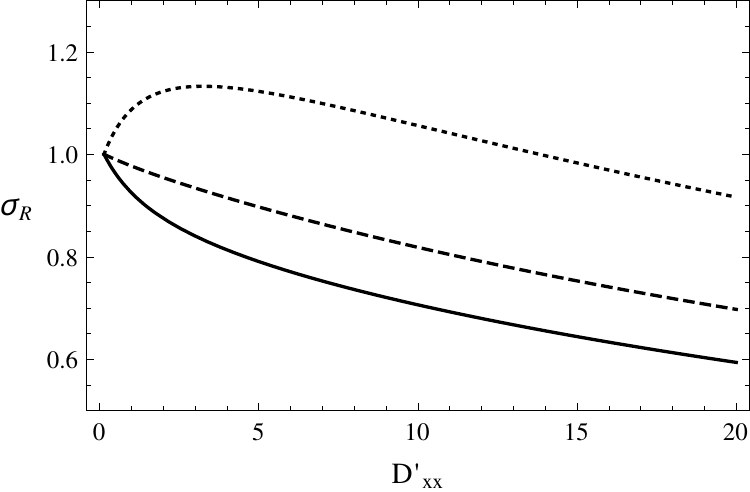}
  %\caption{a)}
\end{subfigure}\\
\begin{subfigure}{.5\textwidth}
  \centering
  \includegraphics[width=0.9\linewidth]{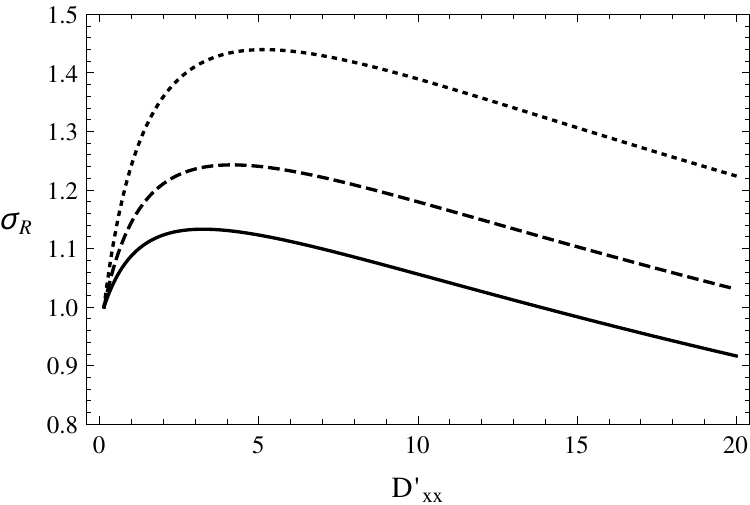}
  %\caption{b)}
\end{subfigure}
\caption{Renormalized entropy production $\sigma_R$ as a function of $D'_{xx}$ according to Eq. \eqref{sigmaR}. 
We set $\gamma'=1$, $D'_{pp}=2$, and $D'_{px}=0.2$  for all curves. Left panel: $c'_4(0)=0.5$ and $c'_5(0)=1$ (solid); $c'_4(0)=1$ and $c'_5(0)=1$ (dashed); 
$c'_4(0)=1$ and $c'_5(0)=0.5$ (dotted). Right panel: $c'_4(0)=1$ and $c'_5(0)=0.5$ (solid); $c'_4(0)=1$ and $c'_5(0)=0.4$ (dashed); 
$c'_4(0)=1$ and $c'_5(0)=0.3$ (dotted). In both figures the curves start at the minimum allowed value in accordance with \eqref{Dekkerineq}.} \label{fig56}
\end{figure}
%%%%%%%%%%%%%%%%%%%%%%%%%%%%%%%%%%%%%%%%%%%%%%%%%%%%%%%%%%%%%%%%%%%%%%%%%%%%%%%%%%

We remind the reader that we are only interested in $D'_{xx} \in [D'^{\text{min}}_{xx}, \infty)$, therefore we 
introduce a renormalized entropy production
\begin{equation}
 \sigma_R \big(D'_{xx}, c'_4(0), c'_5(0) \big)=\frac{\sigma\big(D'_{xx}\big)}{\sigma\big(D'^{\text{min}}_{xx}\big)}, \label{sigmaR}
\end{equation}
which has the following property
\begin{eqnarray}
 &&\frac{\sigma_R \big(D'_{xx}, c'_4(0), c'_5(0) \big)}{\sigma_R \big(D'_{xx}, c'_4(0), c'_5(0)=0 \big)}= 1+\Big[
 16 D'_{pp} c'^2_5(0) \Big.
\nonumber\\ 
 && \Big. +8 D'_{px} c'_4(0) c'_5(0) +32 \gamma'D'_{px} c'^2_5(0) \Big] \times \Big[\gamma'^2+4D'^2_{px} \Big.
 \nonumber \\
&& \Big. -4 D'_{pp}D'_{xx}\Big] 
\Big/  
 \Big[ D'_{xx}\Big(\gamma'^2 \big(c'_4(0)+4\gamma' c'_5(0)\big)^2  \Big.\Big.\nonumber\\
 && \Big. \Big. 
 + 4 \big (4 D'_{pp} c'_5(0)+D'_{px} c'_4(0)+\gamma'D'_{px} c'_5(0) \big)^2\Big) \Big]\leqslant1, \nonumber
\end{eqnarray}
because $\gamma'^2+4D'^2_{px}-4 D'_{pp}D'_{xx}\leqslant 0$ (see Eq. \eqref{Dekkerineq}) and $D'_{px}/D'_{pp} \ll 1$ (see Eq. \eqref{derivedcoeff}). 
Figures in \ref{fig56} show indeed that smaller and positive $c'_5(0)$-s yield
larger $\sigma_R$-s. Through the definition of this new quantity it is guaranteed that
the place of the global maximum is always at the same, yet unknown, value $D'^{\text{?}}_{xx}$ and 
\begin{equation}
 \max_{c'_4(0), c'_5(0)} \sigma_R \big(D'_{xx}, c'_4(0), c'_5(0) \big)=
 \frac{ D'_{xx} }{4\bar n^{\text{st}}+2}
 \log \left(\frac{\bar n^{\text{st}}+1}{\bar n^{\text{st}}}\right)\frac{1}{C}, \nonumber
\end{equation}
where
\begin{equation}
 C=\lim_{D'_{xx} \to D'^{\text{min}}_{xx}}  \frac{ D'_{xx} }{4\bar n^{\text{st}}+2}
 \log \left(\frac{\bar n^{\text{st}}+1}{\bar n^{\text{st}}}\right). \nonumber
\end{equation}
Using the definition of the inverse hyperbolic cotangent function in terms of logarithms
\begin{equation}
 \operatorname{arcoth} \big(x\big) =  \frac12\log\left(\frac{x+1}{x-1}\right)\quad \text{for} \quad x \in (-\infty,-1) \cup (1, \infty), \nonumber
\end{equation}
we arrive at
\begin{eqnarray}
&& \max_{c'_4(0), c'_5(0)} \sigma_R \big(D'_{xx}, c'_4(0), c'_5(0) \big)= \nonumber\\
&& =  \frac{D'_{xx}}{C \big(2\bar n^{\text{st}}+1\big)} \operatorname{arcoth} \big(2\bar n^{\text{st}}+1 \big). \label{todosolve}
\end{eqnarray}
This is a valid transformation because $2\bar n^{\text{st}}+1$ on the interval $[D'^{\text{min}}_{xx}, \infty)$ is strictly larger than one,
i.e., the solution of $\bar n^{\text{st}}=0$ for $D'_{xx}$ yields the roots in Eq. \eqref{rootD}, which are not in the interval
$[D'^{\text{min}}_{xx}, \infty)$.

Let us rewrite the maximum of renormalized entropy production in (\ref{todosolve}) as
\begin{equation}
f(\gamma',T',D'_{xx})=\frac{D'_{xx}}{\big(2\bar n^{\text{st}}+1\big)} \operatorname{arcoth} 
\big(2\bar n^{\text{st}}+1 \big), \label{fest}
\end{equation}
where $T'=k_BT/\omega$ and we have neglected the $D'_{xx}$ independent $C$. We search for the only global maximum of $f(\gamma',T',D'_{xx})$ 
and its position $D'^{?}_{xx}$ in the interval $[D'^{\text{min}}_{xx},\infty)$.  To achieve this goal, 
we need to solve the following equation within the required range:
\begin{equation}
\frac{\partial{f(\gamma',T',D'_{xx})}}{\partial{D'_{xx}}} \bigg|_{D'_{xx}=D'^{?}_{xx}} =0.
\label{trans}
\end{equation}
We observe that $\bar n^{\text{st}}$ is formally equal to $\sqrt{T'+y}-1/2$ with $y\geqslant0$. But, according to the derivation of the master equation we have $k_BT \gg \omega$, i.e., 
$T'\gg 1$ , which
results $\bar n^{\text{st}}\gg1$. Therefore,
\begin{equation}
f(\gamma',T',D'_{xx})\approx \frac{D'_{xx}}{\big(2\bar n^{\text{st}}+1\big)^2}
\end{equation}
and \eqref{trans} yields the following equation:
\begin{equation}
 16 \gamma'^2 T'^2+32 \gamma'^3 T'^2 \frac{\omega}{\Omega}=D'^2_{xx}.
\end{equation}
The physically relevant solution together with the condition $\Omega\gg \omega$ reads
\begin{equation}
 D'^{?}_{xx}=4 \gamma' T' \sqrt{1+2\gamma' \frac{\omega}{\Omega}} \approx 4 \gamma' T'. \label{oursolution1}
\end{equation}

\begin{figure}
\centering
\begin{subfigure}{.5\textwidth}
  \centering
  \includegraphics[width=0.9\linewidth]{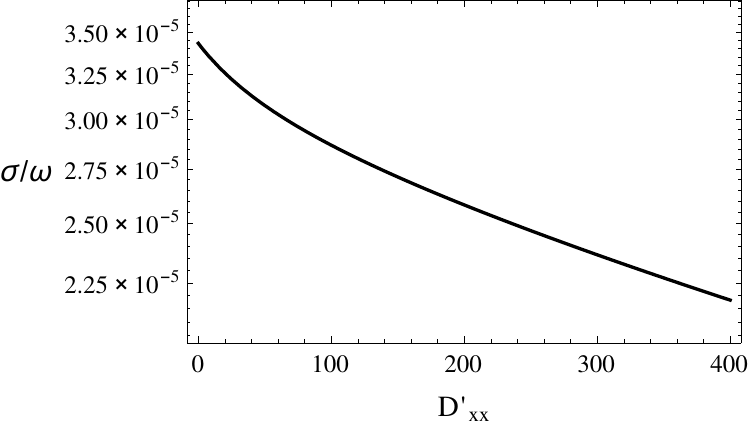}
  %\caption{a)}
\end{subfigure}\\
\begin{subfigure}{.5\textwidth}
  \centering
  \includegraphics[width=0.9\linewidth]{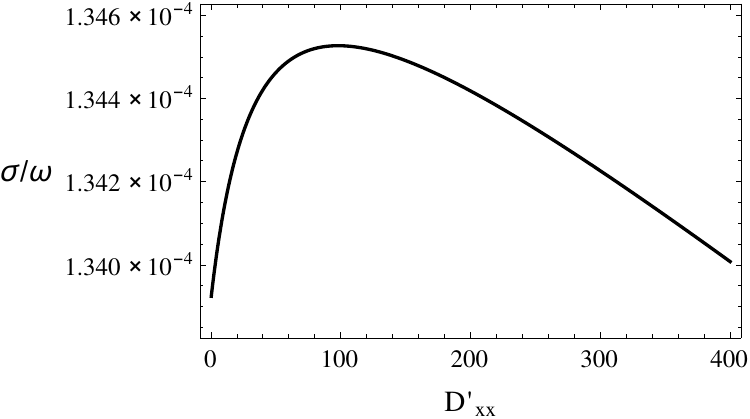}
  %\caption{b)}
\end{subfigure}
\caption{Semilogarithmic plot of the entropy production rate $\sigma/\omega$ as a function of $D'_{xx}$ for the initial condition in \eqref{condspec2}. We set $\omega/\Omega=0.1$, $T'=100$ and
$x=0.1$. Left panel: $\gamma'=1$. Right panel: $\gamma'=4$. In both figures the curves start at the minimum allowed value in accordance with \eqref{Dekkerineq}.} \label{fig78}
\end{figure}

{\it Initial state in the close neighborhood of the steady state.} First, we consider the following type of initial conditions
\begin{eqnarray}
 c'_1(0)&=&c'^{\text{st}}_1+x, \quad c'_2(0)=c'^{\text{st}}_2+x,\quad c'_3(0)=c'^{\text{st}}_3+x, 
 \nonumber \\
  c'_4(0)&=&0,\quad c'_5(0)=0, \label{condspec2}
\end{eqnarray}
where the parameter $x>0$ with $x \ll c'^{\text{st}}_i$ $i\in \{1,2,3\}$ characterizes in a simple way the close vicinity of the steady state. In this situation we have
\begin{equation}
 \lim_{D'_{xx} \to \infty} \frac{\sigma}{\omega}=0,
\end{equation}
which means that $\sigma$ has a global maximum as a function of $D'_{xx}$. We are able to employ again the relation $\bar n^{\text{st}} \gg 1$ to approximate 
$\log \left(\frac{\bar n^{\text{st}}+1}{\bar n^{\text{st}}}\right)$ with $2/(2 \bar n^{\text{st}} +1)$, which 
yields $\sigma$ as a fraction of two polynomials. Now, the position of the global maximum as a function of $\gamma'$ has two separate cases. In order to determine both global maximums we solve
\begin{equation}
\frac{\partial{\sigma}}{\partial{D'_{xx}}} \bigg|_{D'_{xx}=D'^{?}_{xx}} =0,
\label{trans2}
\end{equation}
which results in a third order polynomial equation in $D'_{xx}$. The biggest root of \eqref{trans2} is the position of the global maximum $D'^{?}_{xx}$ and for simplification we have employed 
the $\omega/\Omega \ll 1$ condition, i.e., $\omega/\Omega \approx 0$. For the underdamped $\gamma'<1$, critically damped $\gamma'=1$ and extremely overdamped $\gamma' \gg 1$ situations the 
position of the global maximum $D'^{?}_{xx}=D'^{\text{min}}_{xx}$, because we are allowed to investigate $\sigma$ on the interval $[D'^{\text{min}}_{xx}, \infty)$ due to the condition in 
\eqref{Dekkerineq}. The second 
case constitutes those values of $\gamma'$ when the system is moderately overdamped and then the position of the global maximum $D'^{?}_{xx} \approx \gamma' T'/4$. Fig. \ref{fig78} shows an 
example of these two
cases.

%%%%%%%%%%%%%%%%%%%%%%%%%%%%%%%%%%%%%%%%%%%%%%%%%%%%%%%%%%%%%%%%%%%%%%%%%%%%%%%
%%%%%%%%%%%%%%%%%%%%%%%%%%%%%%%%%%%%%%%%%%%%%%%%%%%%%%%%%%%%%%%%%%%%%%%%%%%%%%%%%%

Secondly, we consider the following type of initial conditions
\begin{eqnarray}
 c'_1(0)&=&c'^{\text{st}}_1+x, \quad c'_2(0)=c'^{\text{st}}_2+x,\quad c'_3(0)=c'^{\text{st}}_3+x, 
 \nonumber\\
 c'_4(0)&=&y,\quad c'_5(0)=y, 
\end{eqnarray}
with $0<x, y \ll c'^{\text{st}}_i$ $i\in \{1,2,3\}$. This more detailed situation however yields again only global maximums and the two cases found for the initial conditions in \eqref{condspec2}.
One may increase 
the number of independent small parameters to five in order to describe the most general vicinity of the steady state, but it reaches the same conclusions as for \eqref{condspec2}. 
Thus, until this point our 
investigations show that initial states which can be reached from the steady state by applying two unitary operations of squeezing and displacing or they are in the close vicinity of the 
steady state result in
global maxima. In the subsequent example we are going to choose a type of initial conditions which do not fall into the previously described sets.

\begin{figure}
\centering
\begin{subfigure}{.5\textwidth}
  \centering
  \includegraphics[width=0.9\linewidth]{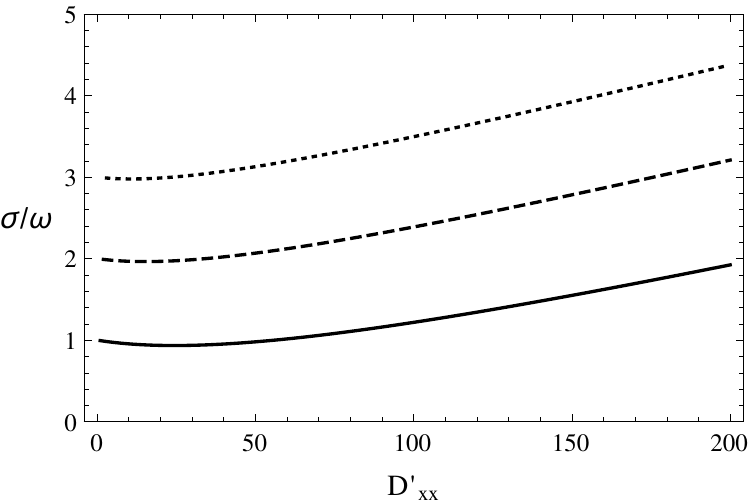}
  %\caption{a)}
\end{subfigure}\\
\begin{subfigure}{.5\textwidth}
  \centering
  \includegraphics[width=0.9\linewidth]{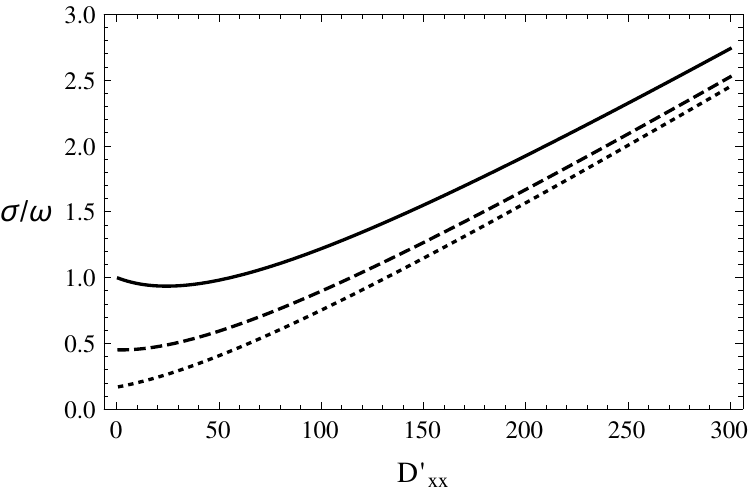}
  %\caption{b)}
\end{subfigure}
\caption{Entropy production rate $\sigma/\omega$ as a function of $D'_{xx}$ for the initial condition in \eqref{condspec3}. We set $\omega/\Omega=0.1$ and
$x=50$. Left panel: $T'=100$; $\gamma'=1$ (solid); $\gamma'=2$ (dashed); $\gamma'=3$ (dotted). Right panel: $\gamma'=1$; $T'=100$ (solid); $T'=125$ (dashed); $T'=150$ (dotted). 
In both figures the curves start at the minimum allowed value in accordance with \eqref{Dekkerineq}.} \label{fig910}
\end{figure}

{\it Initial state has no close relation with the steady state.} Let us consider the initial conditions
\begin{equation}
 c'_1(0)=c'_3(0)=x, \quad c'_2(0)=0,\quad c'_4(0)=0,\quad c'_5(0)=0, \label{condspec3}
\end{equation}
with $x\gg 1$. This choice ensures that $\bar n(0) \gg 1$ and furthermore we have
\begin{equation}
 \lim_{D'_{xx} \to \infty} \frac{\sigma}{\omega}=\infty.
\end{equation}
The above relation shows that we have to search for the global minimum of $\sigma$ on the interval $[D'^{\text{min}}_{xx}, \infty)$. This can be seen in Fig. \ref{fig910}, 
where we have varied $\gamma'$ and $T'$.
It is interesting to note that with the increase of the dimensionless temperature $T'$ the global minimum slides below $D'^{\text{min}}_{xx}$, the minimum allowed value for $D'_{xx}$ 
according to the Dekker 
inequality in \eqref{Dekkerineq}. Thus, for very high temperatures the position of the global minimum $D'^{?}_{xx}=D'^{\text{min}}_{xx}$. In the following we are going to investigate 
the temperature dependence
of $D'^{?}_{xx}$ and determine the condition when position of the global minimum reaches $D'^{\text{min}}_{xx}$.

%%%%%%%%%%%%%%%%%%%%%%%%%%%%%%%%%%%%%%%%%%%%%%%%%%%%%%%%%%%%%%%%%%%%%%%%%%%%%%%
%%%%%%%%%%%%%%%%%%%%%%%%%%%%%%%%%%%%%%%%%%%%%%%%%%%%%%%%%%%%%%%%%%%%%%%%%%%%%%%%%%

Due to $\bar n(0) \gg 1$ we use the simplification
\begin{equation}
 \log \left(\frac{\bar n(0)+1}{\bar n(0)}\right) \approx \frac{2}{2\bar n(0)+1}
\end{equation}
in Eq. \eqref{sigmaan}. This results again in a $\sigma$ which is a fraction of two polynomials. In order to determine the global minimum we solve
\begin{equation}
\frac{\partial{\sigma}}{\partial{D'_{xx}}} \bigg|_{D'_{xx}=D'^{?}_{xx}} =0,
%\label{trans2}
\end{equation}
which yields in a fourth order polynomial equation in $D'_{xx}$. The roots of this polynomial equation together even with the $\omega/\Omega \ll 1$ condition, i.e., $\omega/\Omega \approx 0$, 
are complicated 
functions of $x$, $\gamma'$ and $T'$. Therefore, we introduce $c=x/T'$ and consider the ansatz $D'_{xx}=\gamma' T' y$. Now, the fourth order polynomial simplifies to
\begin{eqnarray}
 &&\left[8 \left(2 \gamma'^2+1\right) y+y^2+16\right]^2-16 c^2 \big[(16 \gamma'^2+1) y^2 \big.
\nonumber\\
 && \big. +32 y+112\big]=0. \label{yeq}
\end{eqnarray}
As $x$ is fixed and we are interested in $D'_{xx}$ (or $y$) as a function of $T'$ and its behavior while $T'$ is increasing. Therefore, we Taylor expand $y$ around $c=0$. It turns out that the 
simplest way to understand the biggest 
root of \eqref{yeq} is to consider the following three cases of the system: extremely underdamped $\gamma' \ll 1$; critically damped $\gamma'=1$; and extremely overdamped $\gamma' \gg 1$. 
These cases result
\begin{equation}
 D'^{?}_{xx}\approx \begin{cases}
-4 \gamma' T' &\text{if } \gamma' \ll 1, \\
-0.68 T'+1.75 c T' +0.46 c^3 T' &\text{if } \gamma'=1, \\
-4 \gamma' T' &\text{if } \gamma' \gg 1. 
\end{cases} \label{strange}
\end{equation}
This shows that between moderately underdamped and overdamped situations $D'^{?}_{xx}$ is a function of the initial condition, i.e., here characterized simply by the parameter $x$. Depending on the 
value of $x$ $D'^{?}_{xx}$ can be larger than the Dekker minimum $D'^{\text{min}}_{xx}$. However, within these type of initial conditions (see Eq. \eqref{condspec3}) the most typical value for 
$D'^{?}_{xx}$ 
is $D'^{\text{min}}_{xx}$ due to \eqref{Dekkerineq}.

\begin{table}[h] \small

		\label{table1}
		\centering
	%	\begin{adjustbox}{width=0.5\textwidth}
		\begin{tabular}{||c||c||}
			
			\hline
			\hline
		  \hline 
			$D_{xx} $  &    General coefficient   \\
			\hline
			$ D^{\text{min}}_{xx} $  &     Minimum coefficient value according \\
			& to Dekker's inequality in Eq. \eqref{Dekkerineq}  \\
			\hline
			$ D'_{xx}$    & General dimensionless coefficient  \\
			\hline
			$D'^{\text{min}}_{xx} $  &   Minimum dimensionless coefficient value  \\
			\hline
			$ D'^{\pm}_{xx} $  & The roots of the quadratic equation in \eqref{mnbv}    \\
			\hline
			$ D'^{?}_{xx} $  &   The dimensionless coefficient value that needs \\ 
			& to be determined from the entropy production  \\
			\hline
			\hline
			\hline
		\end{tabular} 
%	\end{adjustbox}
	\caption{Our conventional symbols for the position diffusion's coefficient used in this section.}  
\end{table}

\section{Discussion and Conclusions}

We have introduced a method to determine the coefficient $D_{xx}$ of the position diffusion term in an extended Caldeira-Leggett master equation. This
term is required to obtain a mathematically consistent master-equation in Lindblad form. The extra of position diffusion term with the coefficient $D_{xx}$ does not affect the mean equations 
of motion for position and momentum, but does introduce an explicit noise source for the equation of the position. It is obvious, that there is no justification for this noise term in the context of 
classical physics, however its presence is due to the mathematical structure of quantum mechanics. The requirement that density matrices are mapped into density matrices validates
the existence of this term. While the modeling of a classical Brown motion requires a L\'evy process whose generator due to the L\'evy-Khintchine formula 
\cite{Sato} can be decomposed into a sum over generators of Poisson, Gaussian and linear drift processes, the quantum analogue of the 
L\'evy-Khintchine formula consists a decomposition with more general generators \cite{Lindblad}. In this context, it should not be surprising that diffusion
terms without classical physical interpretations are present in the quantum theory. The physical origins of such terms may be even in quantum theory very 
puzzling to justify, but in Ref. \cite{Barnett} a theory of quantum Brownian motion is built around a collisional quantum measurement model and
the position diffusion term has been identified as a result of continuous momentum measurements. Thus, one can give a physical interpretation for the 
position diffusion term.  The value of this coefficient is lower bounded
via the Dekker inequality \cite{Dekker2} and obtained by applying a quasi-canonical phase space quantization procedure \cite{Dekker3} or
extending the Caldeira-Leggett model to medium temperatures \cite{Diosi1993}. The latter method heavily depends on the expansion of environment's 
correlation function with respect to the inverse temperature, where quadratic and higher order terms are neglected. In this paper, we have 
studied this coefficient by using the entropy production of the master equation. For simplicity, we have considered an environment 
with Ohmic spectral density and the central particle to be in a harmonic potential. Therefore, the parameters of the model are the temperature 
$T$ of the environment, relaxation rate $\gamma$, mass $m$, and the frequency of the oscillator $\omega$. Then, we have taken an approach, where
the position diffusion's coefficient is unknown in the extended Caldeira-Leggett master equation and investigated the entropy production as a
function of this parameter.

We have used a Gaussian ansatz for the density matrices in the initial conditions, because during the time evolution
the Gaussian form is preserved. We have also shown that these type of states are DSTS states, i.e., Gaussian states \cite{Marian}. Making use of the analytical form of the relative entropy 
between two 
Gaussian states, we have been able to determine analytically the entropy production. In general, the entropy production
is difficult to analyze for all initial conditions, therefore we have restricted our calculation to the following initial states: displaced and squeezed steady states; states in the close 
neighborhood of the 
steady state; and a set of states which do not have any close relation with the steady state. The choice of the initial states resulted in a simplification of the problem, because in this case
the entropy production depends only on the temperature $T$, relaxation rate $\gamma$, mass $m$, frequency of the oscillator $\omega$, the frequency cutoff $\Omega$, 
and the unknown value of the position diffusion's coefficient $D_{xx}$. Then, we have stated that the value of $D_{xx}$ is best defined by
the position of the global extrema of the entropy production, i.e., the fastest or slowest decay of the system towards the steady state. 
In order to determine the position of the entropy production's extrema as a function of $D_{xx}$, we have used the approximations of the high temperature limit. 
We have found three possible values for  position diffusion's coefficient: $D_{xx}\approx2 \gamma k_B T/(m \omega^2)$; the minimum allowed value 
$D_{xx}=\frac{\gamma}{8mk_BT}+\frac{\gamma k_BT}{2m \Omega^2}$ according to the Dekker inequality; and a situation where $D_{xx}$ depends on the initial condition
and decreasing as a polynomial function of $T$, see Eq. \eqref{strange}. First, we comment on the first result, where $D_{xx}$ depends linearly on the temperature. This surprising result has 
been also  reported in other works, see Ref. \cite{Isar}. In this article, $D_{xx}$ is found to be proportional to $1/(2 m \omega) \coth 
[\omega/(2k_B T)]$ ($\hbar=1$) and together with the condition $k_B T /\omega \geqslant 1$ we have in the first-order Taylor series approximation that 
$D_{xx} \sim k_B T/(m \omega^2)$. However, the linear dependence of $D_{xx}$ on the temperature has not been detected yet in typical experiments, see for example 
optomechanical systems \cite{Aspelmeyer} or superconducting tunnel junctions \cite{Kern}. The most convenient result is the minimum value defined by Dekker's inequality, which is small
compared to the coefficient of the momentum diffusion $D_{pp}$. Thus, in the high temperature limit does not affect the mean values and variances of coordinate and momentum, but the master equation
is in the Lindblad form. Finally, the case when $D_{xx}$ depends on the initial condition shows the inadequacy of the entropy production with arbitrary initial states for the task of determining
the coefficient of the position diffusion. However, for certain set of initial conditions this polynomial dependence is replaced by the minimum of the Dekker inequality. In summary, the only situation
which makes sense is the one when we have investigated the entropy production for initial states near the steady state. This argument resembles the typical situation in nonequilibrium statistical 
mechanics,
where the entropy as a function of time is investigated near the equilibrium. 

All known master equations derived from the Caldera-Leggett model \cite{Breuer} are subject to approximations, however they also have to be mathematically consistent. Our work focuses only on a master equation 
in the Markovian limit in order to use the entropy production for determining the coefficient of the position diffusion, which may play an important role in finding evidence for quantum behavior of a mass as 
large as a nanomechanical object \cite{Marshall,BGD}. The presented method gives some partial answer and is still open for further theoretical investigations.

\section*{Acknowledgements}
The authors have profited from helpful discussions with L. Di\'osi, A. Csord\'as, T. Geszti, A. Sauer, F. Weber and G. Alber. This paper is supported by the European Union's
Horizon 2020 research and innovation program under Grant Agreement No. 732894 (FET Proactive HOT–Hybrid Optomechanical Technologies), the National Research 
Development and Innovation Office of Hungary within the Quantum Technology National Excellence Program  (Project No. 2017-1.2.1-NKP-2017-00001), the  National Research 
Development and Innovation Office of Hungary through Grant No.  K120569, and the Hungarian-Czech Joint Research Project MTA/16/05.

\appendix
 \section{Solutions of Eq. \eqref{difftosolve}}
 \label{AppI}

A general solution of \eqref{difftosolve} can be obtained, for example, with the help of the Laplace transformation
\begin{equation}
c_i(z)=\int^{\infty}_0 c_i(t) e^{-zt}\,dt, \quad i \in \{1,2,3,4,5,6\} \nonumber
\end{equation}
which transforms Eq. \eqref{difftosolve} with the initial condition
$c_i(0)$ into an algebraic equation for $c_i(z)$. The first three coefficients $c_1(z)$, $c_2(z)$, and $c_3(z)$ are coupled only to each other
while $c_4(z)$ and $c_5(z)$ form also their own system of linear equations. $c_6(z)$ is simply $c_6(0)/z$. 
Inverting these solution with the help of the inverse relation
\begin{equation}
c_i(t)= \frac{1}{2 \pi i} 
\int_{\mathcal{C}} e^{zt} c_i(z)\, dz, \quad i \in \{1,2,3,4,5,6\} \nonumber
\end{equation}  
we obtain the solution for $c_i(t)$ and the path of integration $\mathcal{C}$ has to 
be chosen in such a way that all poles of $c_i(z)$ are included. The poles of $c_1(z)$, $c_2(z)$, and $c_3(z)$ are: $z=0$,
$z=-2\gamma$, $z=-2 \gamma -2 \sqrt{\gamma^2-\omega^2}$, and $z=-2 \gamma +2 \sqrt{\gamma^2-\omega^2}$. The poles of 
$c_4(z)$, and $c_5(z)$ are: $z=-\gamma -\sqrt{\gamma^2-\omega^2}$ and $z=-\gamma +\sqrt{\gamma^2-\omega^2}$. In the following we 
employ the transformation \eqref{transf}. Furthermore, we consider
the complex parameter $\Omega=\sqrt{\gamma'^2-1}$ and dimensionless time $\tau=\omega t$, which allow us to present the solutions 
in a more concise form. The solutions to the 
first three transformed coefficients are:
\begin{eqnarray}
\!\!&\!\!&\!\!c'_1(\tau)=\frac{4D'_{pp}+D'_{xx} (4\gamma'^2+1)+8\gamma' D'_{px}}{4\gamma'} +\frac{e^{-2 \gamma' \tau}}{4 \gamma' \Omega^2}\times \nonumber\\
\!\!&\!\!&\!\! \times\!\!\Big[4 D'_{pp}+D'_{xx}+4\gamma' D'_{px}-2 \gamma' \big(c'_1(0) + 
2 \gamma' c'_2(0)+ 4 c'_3(0)\big)\Big]
\nonumber \\
&&+\sum_{y=\pm \Omega} 
\Big[ \frac{-4 D'_{pp}+D'_{xx} \big(1-2\gamma'^2 + 2 \gamma' y \big) + 2 c'_1(0) \big(\gamma'-y\big)}
{8 (\gamma'+y) y^2} \Big.
\nonumber\\
&&\Big. +\frac{2 c'_3(0) \big(\gamma'+y\big) + c'_2(0)-D'_{px} \big(\gamma'-y\big) }
{2 (\gamma'+y) y^2}\Big]e^{-2(\gamma' + y)\tau}, \nonumber
\end{eqnarray}
\begin{eqnarray}
&&c'_2(\tau)=-\frac{D'_{xx}}{2} + \frac{e^{-2 \gamma' \tau}}{4 \Omega^2} \Big[-4 D'_{pp} -D'_{xx}+ \Big. \nonumber\\
&&\Big.
2 \gamma'\big(c'_1(0) + 2 \gamma' c'_2(0)+ 4 c'_3(0)-2D'_{px} \big) \Big]+ 
 \nonumber \\
&& \sum_{y=\pm \Omega} e^{-2(\gamma' + y)\tau} \Big[ 
\frac{c'_3(0) \big(1-2 \gamma'^2-2\gamma'y \big)}
{(\gamma'+y) y^2} + \Big( 4 D'_{pp}(\gamma'+y)\Big.\Big. \nonumber \\
&&\Big.
 + D'_{xx} \big(\gamma'-y\big)+4D'_{px}-2 c'_1(0) 
-4c'_2(0)\big(\gamma'+y\big)\Big) \nonumber \\
&&\Big/\Big(8 (\gamma' + y ) y^2 \Big)\Big], \nonumber
\end{eqnarray}
\begin{eqnarray}
&&c'_3(\tau)= \frac{4 D'_{pp}+D'_{xx}}{16 \gamma'}+
\frac{e^{-2 \gamma' \tau}}{16 \gamma' \Omega^2} \Big[4 D'_{pp} + D'_{xx} \Big. \nonumber \\
&& \Big. -2 \gamma'\big(c'_1(0) + 2 \gamma' c'_2(0)+ 4 c'_3(0)-2D'_{px}\big)\Big]
\nonumber \\
&& +\sum_{y=\pm \Omega} e^{-2(\gamma' + y) \tau} \Big[
\Big(4 D'_{pp}\big(1-2\gamma'^2-2 \gamma' y\big) 
-D'_{xx} \nonumber\\
&&+2 c'_1(0) \big(\gamma'+y \big)\Big)\Big/
\Big(32 (\gamma'+y) y^2\Big) -\Big( D'_{px} (\gamma'+y)\nonumber \\
&& +c'_2(0) \big(1-2\gamma'^2-2 \gamma' y\big)+2c'_3(0) \big(3 \gamma'+ y-4\gamma'^3-4\gamma'^2 y \big) \Big) \nonumber\\
&& \Big/
\Big(8 (\gamma'+y) y^2 \big)\Big], \nonumber
\end{eqnarray}
where for $\tau \to \infty$ we obtain the stationary values
\begin{eqnarray}
 c'^{\text{st}}_1&=&\frac{4D'_{pp}+D'_{xx} (4\gamma'^2+1)+8\gamma' D'_{px}}{4\gamma'},\quad c'^{\text{st}}_2=-\frac{D'_{xx}}{2}, \nonumber \\
 c'^{\text{st}}_3&=&  \frac{4 D'_{pp}+D'_{xx}}{16 \gamma'}. \label{stacsol}
\end{eqnarray}
The solutions to the other transformed coefficients are:
\begin{eqnarray}
c'_4(\tau)&=&e^{-(\gamma'+\Omega)\tau} \frac{c'_4(0) \big(\Omega-\gamma'\big) -2 c'_5(0)}{2\Omega} \nonumber \\
&&+
e^{-(\gamma'-\Omega)\tau} \frac{ c'_4(0) \big(\gamma'+\Omega\big)+2 c'_5(0)}{2\Omega}, \nonumber \\
c'_5(\tau)&=&e^{-(\gamma'+\Omega)\tau} \frac{2 c'_5(0)\big(\gamma'+ \Omega\big)+c'_4(0) }{4 \Omega} \nonumber \\
&&+ e^{-(\gamma'-\Omega)\tau} \frac{2 c'_5(0) \big(\Omega-\gamma'\big)-c'_4(0)}{4 \Omega}, \nonumber \\
c'_6(\tau)&=&c'_6(0), \nonumber
\end{eqnarray}
where for $\tau \to \infty$ both $c'_4(\tau)$ and $c'_5(\tau)$ tend to the same stationary value, zero.

In the next step we transform all
$c$'s back to the  $A,B, \dots$ coefficients in the position representation \eqref{xyform} with the help of 
the following relation 
\begin{equation}
\rho(k,\Delta,t)=\int e^{ikx} \rho\left(x+\frac{\Delta}{2}, x-\frac{\Delta}{2}\right) dx. \nonumber
\end{equation}
Evaluating the above integral for the ansatz in the 
position representation (see Eq. \eqref{xyform}) we obtain the relations \cite{Joos}
\begin{eqnarray}
A&=&\frac{1}{x^2_0} \left( c'_3-\frac{c'^2_2}{4c'_1} \right), \quad B=-\frac{1}{x^2_0} \frac{c'_2}{4c'_1}, \quad 
C=\frac{1}{x^2_0}\frac{1}{16 c'_1}, \nonumber \\
D&=&-\frac{1}{x_0} \left( \frac{c'_2c'_4}{2c'_1}-c'_5 \right), \quad E=\frac{1}{x_0}\frac{c'_4}{4c'_1}, \nonumber \\
e^{-c'_6}&=&\sqrt{\frac{\pi}{4C}} \exp \Big\{\frac{E^2}{4C}-N\Big\}=\mathrm{Tr}\{\hat{\rho}\}=1. \label{BtoS}
\end{eqnarray}
These relation are very crucial for our study, because in the subsequent section it will demonstrated that the eigenvalues
and the eigenvectors of the density matrix $\hat{\rho}$ can be constructed with the help of the real parameters
$A(\tau)$, $B(\tau)$, $C(\tau)$, $D(\tau)$, $E(\tau)$ and $N(\tau)$.

\section{Eigenvalues and eigenvectors}
\label{AppII}

The object of this section is to find the eigenvalues and eigenvectors of the Gaussian density matrix in Eq. \eqref{xyform}
\begin{eqnarray}
\rho(x,y)&=&\exp\{-A  \left( x-y \right)^2-iB  \left( x-y \right)  \left( x+y \right) \nonumber \\
&&- C\left( x+y \right) ^{2}
-iD (x-y)-E(x+y) -N\}, \nonumber
\end{eqnarray}
with
\begin{equation}
e^{-N}=\sqrt{\frac{4C}{\pi}} e^{-\frac{E^2}{4C}}. \nonumber 
\end{equation}

In order to solve this problem we introduce the generalized Hermite polynomials $H_n(x,a)$ (see e.g. Ref. \cite{Roman}), which 
are defined by their generating function
\begin{equation}\label{genfun}
\exp\{2xu-au^2\}=\sum_{n=0}^\infty H_n(x,a)\frac{u^n}{n!},
\end{equation}
for all $a>0$. These functions are related to the usual Hermite polynomials as
$H_n(x,a)=a^{n/2}H_n(a^{-1/2}x)$, so $H_n(x)=H_n(x,1)$. The eigenproblem
\begin{equation}\label{rhoeigen}
\int_{-\infty}^{+\infty}\rho(x,y)\phi_n(y) \,dy=\epsilon_n\phi_n(x)
\end{equation}
is solved by
\begin{eqnarray}
&&\phi_n(x)= H_n\left(x+\frac{E}{4C},\frac{1}{4 \sqrt{AC}}\right) \times \nonumber \\
&& \exp\left\{-x^2\left(2\sqrt{AC}+ iB\right)-x\left(\sqrt{A/C}\,E+iD\right)\right\}\!\!, \nonumber \\
&&\epsilon_n=\epsilon_0\epsilon^n, \quad \epsilon_0=\frac{2\sqrt{C}}{\sqrt{A}+\sqrt{C}}, \quad \epsilon=\frac{\sqrt{A}-\sqrt{C}}{\sqrt{A}+\sqrt{C}}. \nonumber
\end{eqnarray}

First, we observe that $A \geqslant C$, otherwise the density matrix has negative eigenvalues. Furthermore, $C >0$ in order to avoid that 
all eigenvalues are equal to zero. The other three parameters $B$, $D$, and $E$ can be any real numbers. These parameter ranges guarantee that
the Gaussian functions of form \eqref{xyform} are states of a quantum harmonic oscillator. In order that the eigenfunctions $\phi_n(x)$ 
form an orthonormal basis, we divide each of them with the square root of the norm resulting a prefactor of \\ $\sqrt{(16AC)^{\frac{2n+1}{4}}/
(\sqrt{\pi} 2^n n!)}$. The obtained formulas extend the result of 
Ref. \cite{joos1985}.

In the following we present a detailed derivation of Eq. \eqref{rhoeigen}. We replace $x$ by $y$ in Eq. \eqref{genfun} and then 
shift $y$ by $y + \kappa$. Multiplying this transformed equality by $\rho(x,y)\exp\{zy^2+wy\}$ and integrating with respect to $y$, we get
\begin{eqnarray}
&&\int_{-\infty}^{+\infty}\exp\{2(y+\kappa)u-au^2\}\rho(x,y)\exp\{zy^2+wy\}\,dy= \nonumber \\
&&=\sum_{n=0}^\infty \frac{u^n}{n!} \int_{-\infty}^{+\infty}\!\!\!\!\!\!\! dy\,
\rho(x,y) H_n(y+\kappa,a)\exp\{zy^2+wy\}\,. \label{inteqproof1}
\end{eqnarray}
The left hand side integrates to
\begin{eqnarray}
&&2 \sqrt{\frac{C}{\tau_- -z}} \exp\left\{-\frac{E^2}{4C}+\frac{(\eta_-+w)^2}{4(\tau_--z)}\right\} \times 
\exp \left\{2x u \frac{\delta}{\tau_--z} \right.\nonumber \\
&&\left.+2u \left(\kappa + 
\frac{\eta_-+\omega}{\tau_--z}\right)-u^2\left(a-\frac{1}{\tau_--z} \right) \right\} \times \nonumber \\
&&\exp \left\{x^2 \left(\frac{\delta^2}{\tau_--z}-\tau_+\right)
+x \frac{\delta(\omega+\eta_-)-(\tau_--z) \eta_+}{\tau_--z} \right\},\nonumber\\ \label{expexpanded}
\end{eqnarray}
where
\begin{equation}
\tau_\pm=A+C\pm iB,\quad \eta_\pm=i D\pm E, \quad \delta=A - C. \nonumber
\end{equation}
In order that Eq. \eqref{inteqproof1} is indeed an eigenvalue equation (see Eq. \eqref{rhoeigen}) with the ansatz eigenvector
$\phi_n(y)=H_n(y+\kappa,a)\exp\{zy^2+wy\}$, we require that the coefficients of $x^2$ and $x$ in \eqref{expexpanded} coincide with 
the coefficients of $y^2$ and $y$ in $\exp\{zy^2+wy\}$. Thus, we can determine $z$ and $\omega$ from the following equations:
\begin{equation}
 z=\frac{\delta^2}{\tau_--z}-\tau_+, \quad
 \omega=\frac{\delta(\omega+\eta_-)-(\tau_--z) \eta_+}{\tau_--z}. \nonumber
\end{equation}
We obtain from the quadratic equation of $z$ the solutions $z=-iB \pm 2 \sqrt{AC}$. As both $A$ and $C$ are positive real numbers the solution
$-iB + 2 \sqrt{AC}$ renders the integral in \eqref{inteqproof1} divergent. Therefore, we consider only $z=-iB - 2 \sqrt{AC}$. This yields that
$\omega=-\sqrt{A/C}\, E-i D$ and
\begin{equation}
\exp\left\{-\frac{E^2}{4C}+\frac{(\eta_-+w)^2}{4(\tau_--z)}\right\}=\exp\left\{-\frac{E^2}{4C}+\frac{E^2}{4C}\right\}=1. \nonumber 
\end{equation}
Now, let $u'=u \delta /(\tau_- -z)$, so that the exponent of second exponential factor in \eqref{expexpanded} reads
\begin{equation}
 2\left(x + \frac{\kappa (\tau_- -z)+\eta_-+\omega }{\delta}\right)u'- \frac{a (\tau_- -z)^2-(\tau_- -z)}{\delta^2} u'^2. \nonumber
\end{equation}
The exponent of the above expression has to coincide with the generating function of $H_n(x+\kappa,a)$, which results for $\kappa$ and $a$
the following equations:
\begin{equation}
 \kappa=\frac{\kappa (\tau_- -z)+\eta_-+\omega }{\delta}, \quad a=\frac{a (\tau_- -z)^2-(\tau_- -z)}{\delta^2}. \nonumber
\end{equation}
Substituting the values of $z$ and $\omega$ we get
\begin{equation}
 \kappa=\frac{E}{2C},\quad a=\frac{1}{4\sqrt{AC}}. \nonumber
\end{equation}
In summary, \eqref{expexpanded} reads
\begin{equation}
\frac{2\sqrt{C}}{\sqrt{A}+\sqrt{C}}  e^{2(x+\kappa)u'-au'^2} e^{-x^2 \left(-iB - 2 \sqrt{AC}\right) -x \left(\sqrt{A/C}\, E+i D\right)}. 
\nonumber
\end{equation}
Now, substituting the above expression into \eqref{inteqproof1} and rewriting the generating function in terms of generalized Hermite polynomials,
we get
\begin{eqnarray}
&&\sum_{n=0}^\infty \frac{u^n}{n!}\frac{2\sqrt{C}}{\sqrt{A}+\sqrt{C}} \Big(\frac{\sqrt{A}-\sqrt{C}}{\sqrt{A}+\sqrt{C}}\Big)^n \times
\nonumber\\
&& \times  H_n(x+\kappa,a) e^{-x^2 \left(-iB - 2 \sqrt{AC}\right) -x \left(\sqrt{A/C}\, E+i D\right)}= \nonumber \\
&&=\sum_{n=0}^\infty \frac{u^n}{n!} \int_{-\infty}^{+\infty} \rho(x,y) H_n\big(y+\kappa,a\big) \times \nonumber\\
&& \times e^{-x^2 \left(-iB - 2 \sqrt{AC}\right) -x \left(\sqrt{A/C}\, E+i D\right)}\,dy, \nonumber
\end{eqnarray}
where we have used the relation
\begin{equation}
u'=u \delta /(\tau_- -z)=u\frac{\sqrt{A}-\sqrt{C}}{\sqrt{A}+\sqrt{C}}. \nonumber 
\end{equation}
We obtain the desired result by comparing the coefficients of $u^n$ on each side of the equation.


\begin{thebibliography}{99}
%
\bibitem{CLmodel} A. O. Caldeira and A. J. Leggett, Physica {\bf 121A}, 587 (1983).
%
\bibitem{Weiss} U. Weiss, {\it Quantum Dissipative Systems} (World Scientific, Singapore, 1999).
%
\bibitem{Grabert} H. Grabert, P. Schramm, and G.-L. Ingold, Phys. Rep. {\bf 168}, 115 (1988).
%
\bibitem{Unruh} W. G. Unruh and W. H. Zurek, Phys. Rev D {\bf 40}, 1071 (1989).
%
\bibitem{Hu} B. L. Hu, J. P. Paz, and Y. Zhang,  Phys. Rev. D {\bf 45}, 2843 (1992).
%
\bibitem{Fleming} C. H. Fleming, A. Roura, B. L. Hu, Ann. Phys. {\bf 326}, 1207 (2011).
%
\bibitem{Gorini} V. Gorini, A. Kossakowski, and E. C. G. Sudarshan, J. Math. Phys. {\bf 17}, 821 (1976).
% 
\bibitem{Lindblad} G. Lindblad, Commun. Math. Phys. {\bf 48}, 119 (1976).
%
\bibitem{Feynman} R. P. Feynman and F. L. Vernon, Ann. Phys. (USA) {\bf 24}, 118 (1963).
%
\bibitem{LajosEu} L. Di\'osi, Europhys. Lett. {\bf 22}, 1 (1993).
%
\bibitem{Halliwell} J. J. Halliwell and A. Zoupas, Phys. Rev. D {\bf 52}, 7294; {\bf 55}, 4697 (1995).
%
\bibitem{Senitzky} I. R. Senitzky, Phys. Rev. {\bf 119}, 670 (1960).
%
\bibitem{Dekker1} H. Dekker, Phys. Rev A {\bf 16}, 2126 (1977).
%
\bibitem{Dekker2} H. Dekker and M. C. Valsakumar, Phys. Lett. {\bf 104A}, 67 (1984).
%
\bibitem{Diosi1993} L. Di\'osi, Physica A {\bf 199}, 517 (1993).
%
\bibitem{Dekker3} H. Dekker, Physica {\bf 95A}, 311 (1979).
%
\bibitem{Spohn} H. Spohn, J. Math. Phys. {\bf 19}, 1227 (1978).
%
\bibitem{Prigogine} I. Prigogine, Science {\bf 201}, 777 (1978).
%
\bibitem{Martyushev} L. M. Martyushev and V. D. Seleznev, Phys. Rep. {\bf 426}, 1 (2006).
%
\bibitem{Marian} P. Marian, T. A. Marian, and H. Scutaru, Phys. Rev. A {\bf 69}, 022104 (2004).
%
\bibitem{PetzOhya} M. Ohya and D. Petz, {\it Quantum Entropy and its Use} (Springer-Verlag, New York, 1993)
%
\bibitem{Breuer} H.-P. Breuer and F. Petruccione {\it The theory of open quantum systems} (Oxford
University Press, Oxford, 2002)
%
\bibitem{SaSc} A. Sandulescu and H. Scutaru, Ann. Phys. (N.Y.), {\bf 173}, 277 (1987).
%
\bibitem{Lindblad2} G. Lindblad, Comm. Math. Phys. {\bf 40}, 147 (1975).
%
\bibitem{Uhlmann} A. Uhlmann, Comm. Math. Phys. {\bf 54}, 21 (1977).
%
\bibitem{Sato} K. Sato, {\it L\'evy Processes and Infinitely Divisible Distributions} (Cambridge
University Press, Cambridge, 1999)
%
\bibitem{Barnett} S. M. Barnett and J. D. Cresser, Phys. Rev. A {\bf 72}, 022107 (2005).
%
\bibitem{Isar} A. Isar, A. Sandulescu, H. Scutaru, E. Stefanescu, and W. Scheid, Int. J. Mod. Phys. E {\bf 3}, 635 (1994).
%
\bibitem{Aspelmeyer} M. Aspelmeyer, T. J. Kippenberg, and F. Marquardt, Rev. Mod. Phys. {\bf 86}, 1391 (2014).
%
\bibitem{Kern} B. J\"ack, J. Senkpiel, M. Etzkorn, J. Ankerhold, C. R. Ast, and K. Kern, Phys. Rev. Lett. {\bf 119}, 147702 (2017).
%
\bibitem{Marshall} W. Marshall, C. Simon, R. Penrose, and D. Bouwmeester, Phys. Rev. Lett. {\bf 91}, 130401 (2003).
%
\bibitem{BGD} J. Z. Bern\'ad, L. Di\'osi, T. Geszti, Phys. Rev. Lett. {\bf 97}, 250404 (2006).
%
\bibitem{Joos} E. Joos, H. D. Zeh, C. Kiefer, D. Giulini, J. Kupsch, and I.-O. Stamatescu
{\it Decoherence and the Appearance of a Classical World in Quantum 
Theory} (Springer-Verlag, Berlin, 1996); Appendix A2.
%
\bibitem{Roman} S. Roman, {\it The umbral calculus} (Academic Press, New York, 1984).
%
\bibitem{joos1985} E. Joos and H. D. Zeh, Z. Phys. B {\bf 59}, 223 (1985).
%
\end{thebibliography}
\end{document}